# Criterion of Quantum Phase Synchronization in Continuous Variable Systems by Local Measurement

Shao-Qiang Ma(马少强), Xiao Zheng(郑晓), Guo-Feng Zhang(张国锋)[*]

*Key Laboratory of Micro-Nano Measurement-Manipulation and Physics (Ministry of Education), School of Physics, Beihang University, Xueyuan Road No. 37, Beijing 100191, China*

**Abstract:** Phase synchronization was proved to be unbounded in quantum level, but the witness of phase synchronization is always expensive in terms of the quantum resource and non-local measurements involved. Based on the quantum uncertainty relation, we construct two local criterions for the phase synchronization in this paper. The local criterions indicate that the phase synchronization in the quantum level can be witnessed only by the local measurements, and the deduction has been verified in the optomechanics system in numerical way. Besides, by analyzing the physical essence of the phase synchronization in quantum level, we show that one can prepare a state, which describes two synchronized oscillators with no entanglement between them. Thus, the entanglement resource is not necessary in the occurrence of the ideal phase synchronization, and also the reason for this phenomenon is discussed.

## I. Introduction

Spontaneous synchronization refers to the phenomenon that the two or more weakly coupled systems with different natural frequencies synchronize their motions only due to their mutual interaction [1-4]. The occurrence of synchronization is first discovered by Huygens in two coupled pendulum clocks, and then has been observed in so many different settings, such as the collective lightning of fireflies, the beating of heart cells and chemical reaction [2]. In classical mechanics, the spontaneous synchronization has been widely studied [5-8], and there exist standard methods to verify whether the motion of two systems is synchronized [2].

In quantum level, the spontaneous synchronization has been considered from different aspects: clock synchronization [9-12], synchronization in oscillator networks [13-21], and synchronization between two atomic ensembles [22]. Notably, due to the absence of the phase space trajectories, the extension of the notion of phase synchronization from classical mechanics to its quantum counterpart is not straightforward [23]. Lots of work has been done and great progress has been made in the research. For instance, the seminal paper [24] proposed a measure for the synchronization, and deduced that the phase synchronization is unbounded in quantum level. However, due to the non-locality of phase difference operator, the detection of phase synchronization in quantum level is always expensive in terms of the quantum resource and non-local measurement involved [25,26].

In this paper, we propose two local criteria for the quantum phase synchronization in the continuous variable systems [27]. The criteria indicate that the phase synchronization in the quantum level can be identified only by local measurements and classical communication. Thus, the criteria save us, especially the experimenters, some quantum resources in the phase synchronization witness. The criteria obtained have been verified by numerical result in optomechanics system [28, 29]. Meanwhile, the entanglement and synchronization are both associated with the correlations between two or more systems, and thus the investigation of relationship between them becomes important [24]. Previous works mainly investigated this relation in numerical way, and deduced that occurrence of the phase synchronization had no relationship with the entanglement. Despite these amazing breakthroughs, the investigation has not stopped, because none of them can give the physical essence behind such a phenomenon, namely the reason why there exists no relationship between entanglement and synchronization. Here, we investigate this relationship in analytical way. By analyzing the physical essence of occurrence of phase synchronization, we show that one can prepare a state, which describes two synchronized oscillators with no entanglement between them. Thus, the ideal phase synchronization can occur without entanglement. Different from previous works, our investigation discusses the essential reason for this phenomenon.

The outline of the paper is as follows. In Sec. II, utilizing the uncertainty relation, we construct two local criteria. In Sec. III, the criteria obtained are verified in the optomechanics system by a numerical simulation. In Sec. IV, the discussion about the relationship between the entanglement and quantum phase synchronization is presented. Finally, Sec. V is devoted to the discussion and conclusion.

## II. Deduction of the local criteria for quantum phase synchronization

Considering two classical continuous variable subsystems $S_1$ and $S_2$ characterized by canonical variables $q_j(t)$ and $p_j(t)$ $(j = 1, 2)$, the classical phase synchronization between them occurs when $\varphi_-(t) = \varphi_1(t) - \varphi_2(t)$ is locked with $\varphi_j(t) = \arctan(p_j(t)/q_j(t))$ [24]. In quantum level, in order to investigate the phase synchronization, we should construct the phase difference operator $\hat{\varphi}_- = \hat{\varphi}_1 - \hat{\varphi}_2$, where $\hat{\varphi}_j$ stands for the phase operator of subsystem $S_j$. In classical mechanics, the investigation of phase synchronization mainly focused on the expected phase, and thus the locked expected phase difference means that the classical phase synchronization occurs. In quantum level, the investigation of phase synchronization mainly focused on the effect of the quantum fluctuation on the classical phase synchronization, and thus we say that the quantum phase synchronization appears when the fluctuation of the phase difference operator can be arbitrarily small, namely $\Delta(\hat{\varphi}_1 - \hat{\varphi}_2)^2 \leq \varepsilon$ with $\Delta\hat{\varphi}_-^2$ being the variance of $\hat{\varphi}_1 - \hat{\varphi}_2$ and $\varepsilon$ being a given precision value [24]. This definition of quantum phase synchronization allows us target the effect of the quantum fluctuation directly and reveal the regimes where synchronization is obtained in the quantum level. Based on the definition of classical phase synchronization and quantum phase synchronization, we can see that the ideal phase synchronization in quantum level appears when the classical phase synchronization and the quantum phase synchronization occur at the same time.

Notably, the detection of classical phase synchronization can be easily achieved, but the detection of $\Delta\hat{\varphi}_-^2 = \langle\delta\hat{\varphi}_1^2\rangle + \langle\delta\hat{\varphi}_2^2\rangle - \langle\delta\hat{\varphi}_1\delta\hat{\varphi}_2 - \delta\hat{\varphi}_2\delta\hat{\varphi}_1\rangle$ is, in general, expensive in terms of the resource involved, especially for the subsystem with long distance [25]. Because the non-local term $\langle\delta\hat{\varphi}_1\delta\hat{\varphi}_2 - \delta\hat{\varphi}_2\delta\hat{\varphi}_1\rangle$ involves the non-local measurement [26], where $\delta\hat{\varphi}_j = \hat{\varphi}_j - \langle\hat{\varphi}_j\rangle$ and $\langle\hat{\varphi}_j\rangle$ is the expectation value of $\hat{\varphi}_j$. In the following, we will introduce the phase operator, and then deduce two local criterions for quantum phase synchronization. The local criterions can be obtained only by local measurements, and thus save us some quantum resources in the phase synchronization witness.

Assume the annihilation operators of the two quantum continuous variable subsystem are $\hat{a}_j = [\hat{q}_j + i\hat{p}_j]/\sqrt{2}$ with $\hat{q}_j$ and $\hat{p}_j$ being quadrature operators of the subsystem and obeying the canonical commutation rules $[\hat{q}_j, \hat{p}_{j'}] = i\delta_{jj'}$. According to Ref. [24], the fluctuation of the phase operator can be interpreted by $\hat{p}'_j$, the anti-Hermitian part of $\hat{a}'_j = [\hat{q}'_j + i\hat{p}'_j]/\sqrt{2}$, where $\hat{a}'_j(t) = e^{-i\langle\hat{\varphi}_j\rangle}\delta\hat{a}_j$ can be obtained by making a rotation transformation on $\delta\hat{a}_j$. The method can only be used to investigate the quantum phase synchronization between the systems with the same amplitude. To investigate more general cases, the method is modified by the Ref. [23] as:

$$\delta\hat{\varphi}_j = \frac{\hat{p}'_j}{\sqrt{2n_j(t)}} = \frac{-\sin\langle\hat{\varphi}_j\rangle\delta\hat{q}_j + \cos\langle\hat{\varphi}_j\rangle\delta\hat{p}_j}{\sqrt{2n_j(t)}} \quad (1)$$

where $\sqrt{n_j(t)}$ is the amplitude of $\langle\hat{a}_j\rangle$.

In fact, there exists no foundational difference between the operator $\delta\hat{\varphi}_j = \hat{p}'_j/\sqrt{2n_j(t)}$ and $\hat{p}'_j$ in interpreting the fluctuation of the phase operator, because the only difference between them is a real coefficient, as shown in Fig.1.

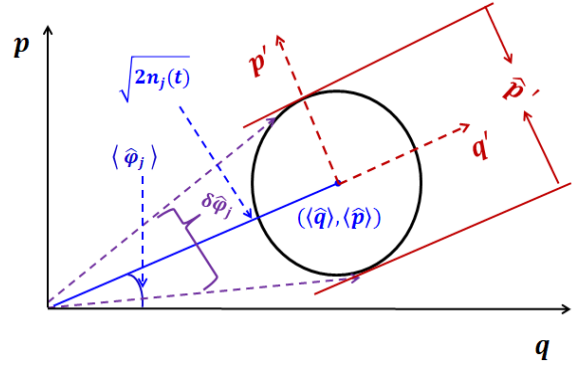

Fig.1. $(q', p')$ is obtained by making a translation and rotation transformation on $(q, p)$ and the angle of rotation is $\langle\hat{\varphi}_j\rangle$. Obviously, the phase fluctuation can be interpreted by $\hat{p}'_j$, and also can be described by the $\hat{p}'_j/\sqrt{2n_j(t)}$.

Denoting $\mathcal{M} = \delta\hat{\varphi}_1 - \delta\hat{\varphi}_2 - \langle\delta\hat{o}\delta\hat{\varphi}_1\rangle\delta\hat{o}/\langle\delta\hat{o}\delta\hat{o}\rangle + \langle\delta\hat{o}\delta\hat{\varphi}_2\rangle\delta\hat{o}/\langle\delta\hat{o}\delta\hat{o}\rangle$, where $\hat{o} = \hat{q}_j$ or $\hat{p}_j$, and taking advantage of $\langle\mathcal{M}\mathcal{M}^\dagger\rangle \geq 0$ [30] and Eq.(1), one can obtain a lower bound for $\Delta(\hat{\varphi}_1 - \hat{\varphi}_2)^2$ (for more details, please see the appendix):

$$\Delta(\hat{\varphi}_1 - \hat{\varphi}_2)^2 \geq \max\{L_{q_1}, L_{p_1}, L_{q_2}, L_{p_2}\}, \quad (2)$$

with

$$L_{\hat{q}_j} = \frac{|\cos\langle\hat{\varphi}_j\rangle|^2 |\langle[\hat{p}_j, \hat{q}_j]\rangle|^2}{8n_j \Delta\hat{q}_j^2}, \quad (3a)$$

$$L_{\hat{p}_j} = \frac{|\sin\langle\hat{\varphi}_j\rangle|^2 |\langle[\hat{p}_j, \hat{q}_j]\rangle|^2}{8n_j \Delta\hat{p}_j^2}, \quad (3b)$$

where $[\hat{p}_j, \hat{q}_j] = -i \neq 0$. Obviously, the lower bound tends to zero when the quantum phase synchronization appears. In other words, the quantum synchronization does not occur, namely $\Delta(\hat{\varphi}_1 - \hat{\varphi}_2)^2 > \varepsilon$, when lower bound (2) is greater than $\varepsilon$, i.e. $\max\{L_{q_1}, L_{p_1}, L_{q_2}, L_{p_2}\} > \varepsilon$, and thus the arbitrarily small lower bound (2) is the necessary condition for the quantum phase synchronization. Meanwhile, it can be seen that the lower bound only involves local measurement, and thus we name the lower bound as the local necessary criterion.

Besides, due to the incompatibility between $\hat{q}_j$ and $\hat{p}_j$, the numerators of the lower bound will never be zero at same time, which means the lower bound tends to zero only when $\Delta p_j^2$ and

$\Delta q_j{}^2$ tend to infinite at the same time [31]. In other words, due to $[\hat{p}_j, \hat{q}_j] \neq 0$, the variance of local operator $p_j$ and $q_j$ will inevitably tend to infinite when the quantum phase synchronization occurs.

Then, another local criterion is presented [32] (for more details, please see the appendix):

$$\Delta(\hat{\varphi}_1 - \hat{\varphi}_2)^2 \leq \left(\sum_{i=1}^{2} \frac{1}{2n_i} (\cos(\varphi_i)^2 \langle \delta p_i \delta p_i \rangle - \sin(\varphi_i)\cos(\varphi_i)\langle\{\delta q_i, \delta p_i\}\rangle + \sin(\varphi_i)^2 \langle \delta q_i \delta q_i \rangle)\right)^2 \quad (4)$$

Obviously, the quantum phase synchronization appears, namely $\Delta(\hat{\varphi}_1 - \hat{\varphi}_2)^2 \leq \varepsilon$, when the upper bound (4) is less than $\varepsilon$, which means the small upper bound (4) is the sufficient condition for the occurrence of quantum phase synchronization. Meanwhile, the bound (4) only involves local measurement, and thus is named as the local sufficient criterion.

Based on the discussion above, we can see that the local necessary criterion and the local sufficient criterion can be used to judge whether the quantum phase synchronization occurs only by the local measurement and classical communication, which can save physical resource for the experimenters in the quantum phase synchronization witness.

### III. Optomechanical Systems as an illustration

Optomechanical system (OMS), a promising platform to investigate the synchronization in quantum level, will be used as an illustration to demonstrate the theoretical conclusions obtained above.

#### A. The Hamiltonian of the System

As shown in Fig. 2, the phase synchronization between two membranes oscillators (MO) in an optical cavity will be considered. The Hamiltonian of the system is written as [23, 29]:

$$H = \hbar\omega_c \hat{a}^\dagger \hat{a} + \sum_{j=1}^{2} \hbar\omega_j \hat{b}_j^\dagger \hat{b}_j - \sum_{j=1}^{2} g_j \hat{a}^\dagger \hat{a}(\hat{b}_j^\dagger + \hat{b}_j) + i\hbar(\eta \hat{a}^\dagger e^{-i\omega_L t} - \eta^* \hat{a} e^{i\omega_L t}) \quad (5)$$

where $\omega_c$ and $\omega_j$ stand for the optical and $j$-th mechanical frequency, $\hat{a}$ and $\hat{b}_j$ are the annihilation operator corresponding to them, $g_j$ is the membrane-cavity coupling strength. The first two terms of the Hamiltonian describe free Hamiltonian of the system, the third term is the optomechanical interaction, and the last term describe the input driving by a laser with frequency $\omega_L$ and amplitude $\eta$.

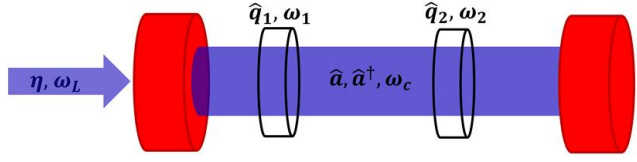

Fig.2: Schematic view of the OMS containing two membranes oscillators in the optical cavity which is pumped with a laser.

The corresponding Heisenberg-Langevin (HL) equations are obtained as: (in the interaction picture with respect to $\hbar\omega_L \hat{a}^\dagger \hat{a}$):

$$\dot{\hat{a}} = \left(i\Delta - \kappa + \sqrt{2}\sum_{j=1}^{2} g_j \hat{q}_j\right)\hat{a} + \eta + \sqrt{2\kappa}\hat{a}^{in},$$

$$\dot{\hat{p}}_j = -\omega_j \hat{q}_j + \sqrt{2}g_j \hat{a}^\dagger \hat{a} - \gamma_j \hat{p}_j + \hat{\xi}_j,$$

$$\dot{\hat{q}}_j = \omega_j \hat{p}_j, \tag{6}$$

where $\Delta = \omega_L - \omega_c$ denotes the detuning of the driving laser from the cavity frequency, $\gamma_j$ is the mechanical damping rate of the $j$-th MO, $\kappa$ is the decay rate of the cavity and $\hat{q}_j = (\hat{b}_j^\dagger + \hat{b}_j)/\sqrt{2}$ as well as $\hat{p}_j = (\hat{b}_j - \hat{b}_j^\dagger)/i\sqrt{2}$ stands for dimensionless position and momentum operators of the $j$-th MO. $\hat{a}^{in}$ stands for the vacuum optical input noise and satisfies the Markovian correlation functions [23]. Assuming each mechanical mode is coupled to a thermal bath at zero temperature and is subject to a Brownian stochastic force $\hat{\xi}_j(t)$. In the limit of high mechanical quality factor, the noise operator $\hat{\xi}_j(t)$ is delta-correlated [33, 34], and the corresponding symmetrized correlation function become $\langle\hat{\xi}_j(t)\hat{\xi}_j(t') + \hat{\xi}_j(t')\hat{\xi}_j(t)\rangle/2 = \gamma_j \delta(t - t')$ [35-37]. The HL equation (6) and the correlations functions can fully describe the dynamics of the system under consideration [23].

## B. Quantum dynamics of the System

We only focus on the quantum phase synchronization, which can be investigated through the fluctuations of the operators around the time-independent mean values. Based on Refs. [23,24], the corresponding dynamical HL equations can be expressed as:

$$\dot{u}(t) = A(t)u(t) + n(t), \tag{7}$$

where the $u(t) = (\delta q_1, \delta p_1, \delta q_2, \delta p_2, \delta X, \delta Y)^T$ is the vector of fluctuation operators with $\delta O = \hat{O} - \langle\hat{O}\rangle$ and the corresponding noises vector is $n(t) = (0, \hat{\xi}_1(t), 0, \hat{\xi}_2(t), \sqrt{\kappa}X^{in}(t), \sqrt{\kappa}Y^{in}(t))^T$. Here the definition of the optical mode quadrature $\delta X = (\delta a + \delta a^\dagger)/\sqrt{2}$ and $\delta Y = (\delta a - \delta a^\dagger)/i\sqrt{2}$ as well as the corresponding noise operator $X^{in} = (a^{in} + a^{in\dagger})/\sqrt{2}$ and $Y^{in} = (a^{in} + a^{in\dagger})/i\sqrt{2}$ are used. Meanwhile, the drift matrix A is given by

$$A = \begin{bmatrix} 0 & \omega_1 & 0 & -u & 0 & 0 \\ -\omega_1 & -\gamma_1 & 0 & 0 & A_1 & B_1 \\ 0 & 0 & 0 & \omega_2 & 0 & 0 \\ 0 & 0 & -\omega_2 & -\gamma_2 & A_2 & B_2 \\ -B_1 & 0 & -B_2 & 0 & -\kappa & M \\ A_1 & 0 & A_2 & 0 & -M & -\kappa \end{bmatrix}, \quad (8)$$

with the elements $A_j = 2g_j \text{Re}(\langle a \rangle)$, $B_j = 2g_j \text{Im}(\langle a \rangle)$ and $M = -\Delta - \sqrt{2} \sum g_j \langle \hat{q}_j \rangle$. Based on the definition of quantum phase synchronization, the dynamics evolution of the quantum phase synchronization can be fully described by the covariance matrix (CM) $V_{ij} = [\langle u_i(t)u_j(t) + u_j(t)u_i(t)\rangle]/2$. The time evolution of the CM is given by [24, 31]:

$$\frac{d}{dt}V(t) = A(t)V(t) + V(t)A^T(t) + D \quad (9)$$

where $D = \text{diag}\{0, \gamma_1, 0, \gamma_2, \kappa, \kappa\}$ is the diffusion matrix. Based on Eq. (9), we can obtain the dynamics evolution of the quantum phase synchronization in numerical way.

### C. Numerical Results

Utilizing Eq. (9), we show that the local necessary criterion and the local sufficient criterion predicted by the previous theoretical analysis can be confirmed by the numerical simulations in OMS, as shown in Fig.3. We can find that the phase synchronization is really bounded by the local necessary criterion and the local sufficient criterion.

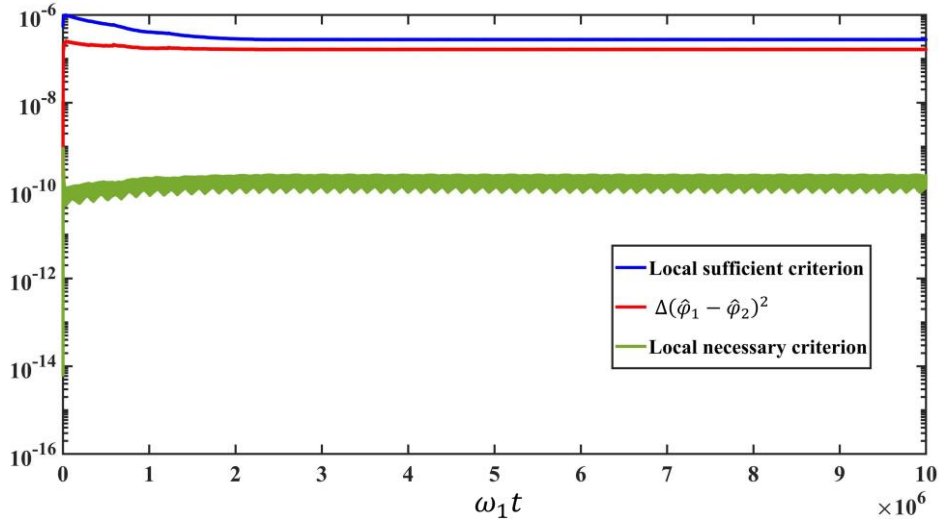

Fig.3: Time evolution of $\Delta(\hat{\varphi}_1 - \hat{\varphi}_2)^2$, local necessary criterion and the local sufficient criterion with respect to the scaled time $\omega_1 t$ for parameters $\eta/\omega_1 = 3600$, $(\omega_1 - \omega_2)/\omega_1 = 0.001, \kappa/\omega_1 = 0.05$, $\Delta/\omega_1 = 1, \gamma_1/\omega_1 = \gamma_2/\omega_1 = 0.000005$ and $\sqrt{2}\, g_1/\omega_1 = \sqrt{2}\, g_2/\omega_1 = 0.00001$. Here the cavity mode is in the vacuum state when t=0, namely $V(0) = \text{diag}\{1/2, 1/2, 1/2, 1/2, 1/2, 1/2\}$ [23].

### IV. Synchronization can occur without Entanglement

The entanglement and the phase synchronization are both associated with the correlations between two or more systems, and therefore, the research on the relationship between them becomes relevant. Ref. [24] concludes that the system can possess the maximum amount of phase synchronization without being necessarily entangled, and the conclusion was verified by Ref. [23] in numerical way. In the following, by introducing the physical essence of phase synchronization in quantum level, we will verify the relationship in analytical way, and provide the reason for such a phenomenon.

We first make a change of the picture on (5) by a rotation unitary transformation $U_r = \exp[r(b_1 b_2^\dagger - b_1^\dagger b_2)]$ with the parameter $r$ being the angle of rotation, and then refer to the picture before and after the transformation as the Schrödinger picture and Single leaking mode picture, respectively. By taking $r = \arctan(g_1/g_2)$, and imposing $|\omega_1 - \omega_2| \ll (g_1^2 + g_2^2)^{3/2}/|g_1 g_2|$, $|(\omega_1 - \omega_2)\sin r \cos r| \ll \kappa$, and $g_1^2 + g_2^2 \ll \omega_1^2 + \omega_2^2$, the Hamiltonian of the system in the Single leaking mode picture becomes [14]:

$$\widetilde{H} = \hbar\omega_c \hat{a}^\dagger \hat{a} + \sum_{j=1}^{2} \hbar\widetilde{\omega}_j \hat{b}_j^\dagger \hat{b}_j - \tilde{g}_2 \hat{a}^\dagger \hat{a}(\hat{b}_2^\dagger + \hat{b}_2) + i\hbar(\eta \hat{a}^\dagger e^{-i\omega_L t} - \eta^* \hat{a} e^{i\omega_L t}) \quad (10)$$

where $\widetilde{\omega}_1 = \omega_1 \cos^2 r + \omega_2 \sin^2 r$, $\widetilde{\omega}_2 = \omega_1 \sin^2 r + \omega_2 \cos^2 r$, $\tilde{g}_2 = g_1 \sin r + g_2 \cos r$ [14]. In the Schrödinger picture the Hamiltonian describes a system where the two MOs are coupled to an optical cavity, and in the Single leaking mode picture, the Hamiltonian describes a situation where only mode 2 is directly coupled to the optical cavity and there exists no interaction between two modes. Consider that the optical cavity is a dissipative cavity and $\eta$ is equal to zero [38]. In the Single leaking mode picture, the mode 2 coupled to the dissipative cavity will be driven to ground state (this is the reason why the new picture is called Single leaking mode picture by Ref. [14]), and the mode 1 is unitary evolution with the frequency $\widetilde{\omega}_1$. In the Schrödinger picture, the state indicates that the whole system, so do the two MOs, oscillates at frequency $\widetilde{\omega}_1$, namely the average phase synchronization occurs [14].

Then, we will show that the ideal phase synchronization can occur without necessary entanglement. Based on the analysis above, the phase synchronization in the Schrödinger picture appears when the mode 2 in the Single leaking mode picture reaches the ground state. Here, we denote the state in the Schrödinger picture as $\rho_s$, and the state in the single leaking mode picture as $\rho_s'$, when the phase synchronization appears. Obviously, there exists no entanglement between mode1 and mode 2 in the Single leaking mode picture when the mode 2 reaches the ground state. That is to say, the entanglement of $\rho_s'$ is equal to zero. Meanwhile, the entanglement of the state

$\rho_s = U_r^\dagger \rho_s' U_r$ cannot be guaranteed to be greater than zero when the entanglement of $\rho_s'$ is equal to zero. For instance, assume that the initial state of the two MOs (in the Schrödinger picture) is prepared as the entangled coherent state:

$$|\varphi(0)\rangle = \sin(\theta)|\alpha\rangle_1|\alpha^*\rangle_2 + \cos(\theta)|0\rangle_1|0\rangle_2 \tag{11}$$

where $\theta \in [0, 2\pi]$, and $|\alpha\rangle_j$ is the coherent state of $j-MO$ with $\alpha = 500\sqrt{2} + i500\sqrt{2}$. Here, we take $g_1 = g_2$, which means $r = \arctan(g_1/g_2) = \pi/4$, and then one has [14]:

$$\rho_s' = \sin(\theta)^2 \left|\tilde{\beta}e^{i\widetilde{\omega_1}t}\right\rangle_1 |0\rangle_2 \langle 0|_2 \left\langle\tilde{\beta}e^{i\widetilde{\omega_1}t}\right|_1 + \cos(\theta)^2 \left|\tilde{\alpha}e^{i\widetilde{\omega_1}t}\right\rangle_1 |0\rangle_2 \langle 0|_2 \left\langle\tilde{\alpha}e^{i\widetilde{\omega_1}t}\right|_1 \tag{12}$$

when the phase synchronization occurs, where $\tilde{\alpha} = 0$, and $\tilde{\beta} = 1000$. Taking advantage of unitary transformation $\rho_s = U_r^\dagger \rho_s' U_r$, one can obtain the synchronized state in the Schrödinger picture:

$$\rho_s = \sin(\theta)^2 \left|500\sqrt{2}e^{i\widetilde{\omega_1}t}\right\rangle_1 \left|-500\sqrt{2}e^{i\widetilde{\omega_1}t}\right\rangle_2 \left\langle-500\sqrt{2}e^{i\widetilde{\omega_1}t}\right|_2 \left\langle 500\sqrt{2}e^{i\widetilde{\omega_1}t}\right|_1$$
$$+ \cos(\theta)^2 |0\rangle_1|0\rangle_2\langle 0|_2\langle 0|_1 \tag{13}$$

Obviously, the entanglement of $\rho_s$ is zero, and the expectation value of $\langle b_1\rangle$ and $\langle b_2\rangle$ on $\rho_s$ can be obtained by straightforward calculation:

$$\langle b_1\rangle(t) = \sin(\theta)^2\, 500\sqrt{2}e^{i\widetilde{\omega_1}t} \tag{14a}$$

$$\langle b_2\rangle(t) = \sin(\theta)^2\, 500\sqrt{2}e^{i(\widetilde{\omega_1}t-\pi)} \tag{14b}$$

It can be seen that both the two MOs oscillate at the same frequency $\widetilde{\omega_1}$, namely the classical phase synchronization between two MOs occurs, and the phase difference between them is $\pi$, which exactly coincide with the numerical result, as shown in Fig.4. Meanwhile, based on (4), we can obtain that the local sufficient criterion of $\rho_s$ is equal to $8*10^{-6}$, which means the quantum phase synchronization appears and the classical phase synchronization will not be destroyed by the quantum fluctuation. Thus, we can obtain that the ideal phase synchronization occurs in the separable state $\rho_s$.

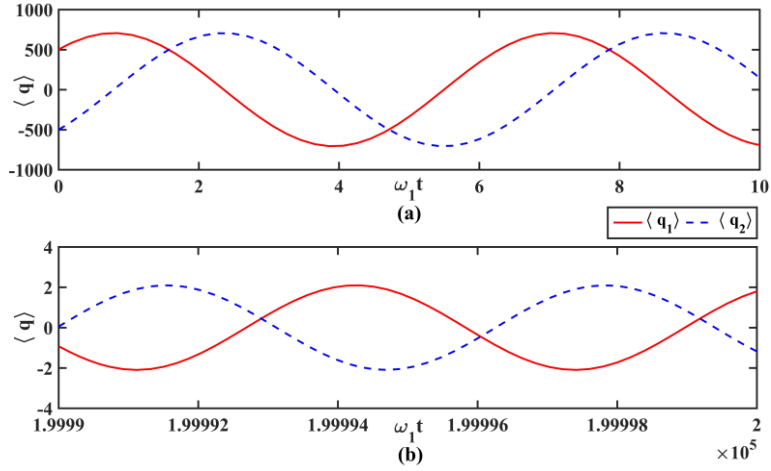

Fig.4: Time evolution of $\langle q_1 \rangle$ and $\langle q_2 \rangle$ with the scaled time $\omega_1 t$, in the initial part of the evolution (a) and in the final part of the evolution (b), for parameters $\eta/\omega_1 = 0$, $\omega_1 - \omega_2/\omega_1 = 0.001$, $\kappa/\omega_1 = 1$, $\Delta/\omega_1 = 0$, $\gamma_1/\omega_1 = \gamma_2/\omega_1 = 0$, $\sqrt{2}\, g_1/\omega_1 = \sqrt{2}\, g_2/\omega_1 = 0.02$, and $\theta = \pi/4$. It can be seen that the phase difference between $\langle q_1 \rangle$ and $\langle q_2 \rangle$ is not equal to $\pi$ in the initial evolution, while in the final evolution the phase difference is equal to $\pi$, as expected by the theory [38].

The discussion above is mainly based on the optomechanics system, but we should mention that the corresponding discussion can be easily extended to other systems. As mentioned above, the reason why we can prepare a separated and synchronized state so as to demonstrate that synchronization can occur without necessary entanglement is that the entanglement of $\rho_s = U_r^\dagger \rho_s' U_r$ cannot be guaranteed to be greater than zero by the unitary transformation $U_r$, when the entanglement of $\rho_s'$ is equal to zero. In fact, the conclusion obtained applies to arbitrary synchronization which is obtained by decaying one mode to ground state in the Single leaking mode picture, because the entanglement of $\rho_s'$ is always equal to zero when one mode is in the ground state. Thus, our deduction obtained can be extended to other systems, where such kind of synchronization can occur, for instance the system of two oscillators coupled to a two-level system [14].

## V. Conclusion

In conclusion, the phase synchronization in the continuous variable system has been investigated in this work. We constructed a local necessary criterion and a local sufficient criterion for the phase synchronization by using the quantum uncertainty relation. In general, the witness of quantum phase synchronization is expensive in in terms of the resource and non-local measurement involved. The local criterions indicate that the quantum phase synchronization can be witnessed only by local measurements, which can save physical resource for us in the quantum phase synchronization detection. The deduction has been verified by numerical results in optomechanics system. Besides, previous work about the relationship between the phase synchronization and the entanglement was mainly done in the numerical way, and found that there exists no foundational relation between them. However, none of them can give the reason for such a phenomenon. Here we investigate this phenomenon from the perspective of physical essence of synchronization, and give the reason that why the entanglement resource is not necessary in the occurrence of the phase synchronization.

**Acknowledgments**

This work is supported by the National Natural Science Foundation of China (Grant No. 11574022).

**Appendix**

Taking

$$\mathcal{M} = \delta\hat{\varphi}_1 - \delta\hat{\varphi}_2 - \frac{\langle \delta\hat{o}\delta\hat{\varphi}_1\rangle\delta\hat{o}}{\langle \delta\hat{o}\delta\hat{o}\rangle} + \frac{\langle \delta\hat{o}\delta\hat{\varphi}_2\rangle\delta\hat{o}}{\langle \delta\hat{o}\delta\hat{o}\rangle}, \tag{A1}$$

and then, we have

$$\langle \mathcal{M}\mathcal{M}^\dagger\rangle = \langle (\delta\hat{\varphi}_- - \frac{\langle \delta\hat{o}\delta\hat{\varphi}_-\rangle\delta\hat{o}}{\langle \delta\hat{o}\delta\hat{o}\rangle})(\delta\hat{\varphi}_- - \frac{\langle \delta\hat{o}\delta\hat{\varphi}_-\rangle\delta\hat{o}}{\langle \delta\hat{o}\delta\hat{o}\rangle})^\dagger\rangle$$

$$= \langle (\delta\hat{\varphi}_- - \frac{\langle \delta\hat{o}\delta\hat{\varphi}_-\rangle\delta\hat{o}}{\langle \delta\hat{o}\delta\hat{o}\rangle})(\delta\hat{\varphi}_- - \frac{\langle \delta\hat{\varphi}_-\delta\hat{o}\rangle\delta\hat{o}}{\langle \delta\hat{o}\delta\hat{o}\rangle})\rangle$$

$$= \langle \delta\hat{\varphi}_-\delta\hat{\varphi}_-\rangle - \frac{|\langle \delta\hat{o}\delta\hat{\varphi}_-\rangle|^2}{\langle \delta\hat{o}\delta\hat{o}\rangle}$$

$$= \Delta(\hat{\varphi}_1 - \hat{\varphi}_2)^2 - \frac{|\langle \delta\hat{o}\delta\hat{\varphi}_-\rangle|^2}{\langle \delta\hat{o}\delta\hat{o}\rangle}. \tag{A2}$$

Based on the Ref. [12] and [30], we have:

$$\langle \mathcal{M}\mathcal{M}^\dagger\rangle \geq 0 \tag{A3}$$

where $\langle \mathcal{M}\mathcal{M}^\dagger\rangle$ is the second-order original moment of the operator $\mathcal{M}$. Taking (A2) into (A3), one have:

$$\Delta(\hat{\varphi}_1 - \hat{\varphi}_2)^2 \geq \frac{|\langle \delta\hat{o}\delta\hat{\varphi}_-\rangle|^2}{\langle \delta\hat{o}\delta\hat{o}\rangle} \geq \frac{|\langle [\hat{o},\hat{\varphi}_-]\rangle|^2}{4\langle \delta\hat{o}\delta\hat{o}\rangle}. \tag{A4}$$

Taking $\hat{o} = \hat{q}_1$, $\hat{p}_1$, $\hat{q}_2$ and $\hat{p}_2$, respectively, and using Eq. (1) in the main text, one can obtain:

$$\Delta(\hat{\varphi}_1 - \hat{\varphi}_2)^2 \geq \frac{|\cos\langle\hat{\varphi}_1\rangle|^2|\langle[\hat{p}_1,\hat{q}_1]\rangle|^2}{8n_1\Delta\hat{q}_1^2};$$

$$\Delta(\hat{\varphi}_1 - \hat{\varphi}_2)^2 \geq \frac{|\sin\langle\hat{\varphi}_1\rangle|^2|\langle[\hat{p}_1,\hat{q}_1]\rangle|^2}{8n_1\Delta\hat{p}_1^2};$$

$$\Delta(\hat{\varphi}_1 - \hat{\varphi}_2)^2 \geq \frac{|\cos\langle\hat{\varphi}_2\rangle|^2|\langle[\hat{p}_2,\hat{q}_2]\rangle|^2}{8n_2\Delta\hat{q}_2^2};$$

$$\Delta(\hat{\varphi}_1 - \hat{\varphi}_2)^2 \geq \frac{|\sin\langle\hat{\varphi}_2\rangle|^2|\langle[\hat{p}_2,\hat{q}_2]\rangle|^2}{8n_2\Delta\hat{p}_2^2}.$$

Thus, the Eqs. (2) and (3) in the main text can be obtained.

Meanwhile, In Ref. [32], A. K. Pati and P. K. Sahu deduced a new uncertainty relation for two incompatible observables A and B, which reads:

$$\Delta(A + B) \leq \Delta A + \Delta B \tag{A5}$$

Taking $A = \hat{\varphi}_1$ and $B = -\hat{\varphi}_2$, one has:

$$\Delta(\hat{\varphi}_1 - \hat{\varphi}_2)^2 \leq (\Delta\hat{\varphi}_1 + \Delta\hat{\varphi}_2)^2$$
$$= \left(\sum_{i=1}^{2} \frac{1}{2n_i}(\cos(\varphi_i)^2 \langle \delta p_i \delta p_i \rangle - \sin(\varphi_i)\cos(\varphi_i)\langle\{\delta q_i, \delta p_i\}\rangle + \sin(\varphi_i)^2 \langle \delta q_i \delta q_i \rangle)\right)^2.$$

Then, we obtain Eq. (4) in the main text.

Thus, in practice, the amplitude of the mode 1 in the Single leaking picture, so does the amplitudes of the two MOs in the Schrodinger picture, will inevitably decrease in long time evolution, as shown in Fig. 4.